\documentclass[universe,communication,accept,pdftex,moreauthors]{Definitions/mdpi} 

\firstpage{1} 
\pubvolume{1}
\issuenum{1}
\articlenumber{0}
\pubyear{2023}
\copyrightyear{2023}
\datereceived{18 May 2023 } 
\daterevised{ 03 July 2023} 
\dateaccepted{04 July 2023 } 
\datepublished{ } 
\hreflink{https://doi.org/} 

\Title{Blinkverse: A Database of Fast Radio Bursts}

\TitleCitation{Blinkverse: A Database of Fast Radio Bursts}

\Author{Jiaying Xu 
 $^{1,}$$^{\dagger}$\orcidA{}, Yi Feng $^{1,}$*$^{,\dagger}$\orcidB{}, Di Li $^{2,1}$*\orcidC{}, Pei Wang $^{2,3}$\orcidD{}, Yongkun Zhang $^{2}$\orcidE{}, Jintao Xie $^{1}$, Huaxi Chen $^{1}$, Han Wang $^{1}$, Zhixuan Kang $^{1}$, Jingjing Hu $^{1}$, Yun Zheng $^{1}$, Chao-Wei Tsai $^{2}$, Xianglei Chen $^{2}$ and Dengke Zhou $^{1}$}

\AuthorNames{Jiaying Xu, Yi Feng, Di Li, Pei Wang, Yongkun Zhang, Jintao Xie, Huaxi Chen, Zhixuan Kang, Jingjing Hu, Yun Zheng, Chao-Wei Tsai, Xianglei and Dengke Zhou}

\AuthorCitation{Xu, J.; Feng, Y.; Li, D.; Wang, P.; Zhang, Y.; Xie, J.; Chen, H.; Kang, Z.; Hu, J.; Zheng, Y.; et al.}

\address{%
$^{1}$ \quad Research Center for Intelligent Computing Platforms, Zhejiang Laboratory, Hangzhou 311100, China; : xujy@zhejianglab.com (J.X.) xiejintao@zhejianglab.com 
 (J.X.); chenhuaxi@zhejianglab.com (H.C.); wanghan@zhejianglab.com (H.W.); kangzx@zhejianglab.com (Z.K.); hujingjing@zhejianglab.com (J.H.); zhengyun@zhejianglab.com (Y.Z.); zdk@zhejianglab.com (D.Z.)\\
$^{2}$ \quad National Astronomical Observatories, Chinese Academy of Sciences, Beijing 100101, China; wangpei@nao.cas.cn (P.W.)
 ykzhang@nao.cas.cn (Y.Z.); cwtsai@nao.cas.cn (C.-W.T.); xiangleichen@nao.cas.cn (X.C.)\\
$^{3}$ \quad Institute for Frontiers in Astronomy and Astrophysics, Beijing Normal University,  Beijing 102206, China; wangpei@nao.cas.cn (P.W.)}

\corres{Correspondence:  yifeng@zhejianglab.com (Y.F.); dili@nao.cas.cn (D.L.) 
}

\firstnote{These authors contributed equally to this work.}

\abstract{The volume of research on fast radio bursts (FRBs) observation have been seeing a dramatic growth. 
	To facilitate the systematic analysis of the FRB population, we established a database platform, Blinkverse 
 (\url{https://blinkverse.alkaidos.cn}), as a central inventory of FRBs from various observatories and with published properties, particularly dynamic spectra from FAST, CHIME, GBT, Arecibo, etc. Blinkverse thus not only forms a superset of FRBCAT, TNS, and CHIME/FRB, but also provides convenient access to thousands of FRB dynamic spectra from FAST, some of which were not available before. Blinkverse is regularly maintained and will be
updated by external users in the future. Data entries of FRBs can be retrieved through parameter searches through FRB location, fluence, etc., and their logical combinations. Interactive visualization was built into the platform. We analyzed the energy distribution, period analysis, and classification of FRBs based on data downloaded from Blinkverse. The energy distributions of repeaters and non-repeaters are found to be distinct from one another.}

\keyword{fast radio bursts; radio astronomy; database} 

\begin{document}

\section{Introduction}

Fast radio burst (FRB) is a type of bright single pulse at radio frequencies, with enormous energy generation of millisecond duration. Among over 700 FRBs that have now been reported since the first discovery in 2007 \citep{lorimer2007}, the majority of FRB discoveries have been found to be one-offs. The number of current repeating sources has reached up to 63. With the rapid increase in the number of FRB discoveries in recent years, a mass of remarkable breakthroughs have been made in the research of FRBs, such as repeaters \citep{spitler2016,chime2019a,chime2019b,li2021}, burst characteristics \citep{chime2020,niu2022}, ambient environment \citep{michilli2018,hilmarsson2021,feng2022a,feng2022b,kirsten2022,anna2022}, and host galaxies \citep{keane2016,chatterjee2017,ravi2019,bannister2019}.

The number of FRB discoveries has increased greatly, which also demands higher requirements for data collation and analysis. Several databases are currently available for a range of FRB properties. 
The Transient Name Server (TNS)  \endnote{\url{https://www.wis-tns.org} 
} is the official IAU mechanism for reporting new astronomical transients (ATs) including FRBs. The Fast Radio Burst Catalogue (FRBCAT)  \endnote{\url{https://frbcat.org} 
} is a specific repository for FRB properties but has not been actively updated since July 2020 \citep{petroff2016}. The contents of FRBCAT have been migrated to the TNS. FRBSTATS  \endnote{\url{https://www.herta-experiment.org/frbstats} 
} provides a platform for recording FRB bursts and a visualization interface to plot the parameter distributions \citep{misirlis2022}. Meanwhile, a clustering method, called density-based spatial clustering of applications with noise (DBSCAN), has been applied in the FRBSTATS platform to distinguish repeaters from non-repeaters automatically. Compared with the available databases mentioned above, our Blinkverse database possesses the most comprehensive information on published FRBs and a dynamic visualization platform for fruitful statistical results. Researchers can obtain the target data by constraining one or more parameter of the FRB. The searching capability of Blinkverse is stronger than previous databases. 

In the following sections, the architecture of the database platform will be introduced in Section \ref{sec2}, including data description and data availability. The advantages of Blinkverse compared to other databases will be subsequently listed in Table \ref{tab2}. Section \ref{sec3} will provide several examples of data analysis using data readily downloaded from our database, such as energy distribution, period analysis, and classification of FRBs.  Concluding remarks are provided in Section \ref{sec4}.

\section{Platform Architecture}\label{sec2}

\begin{figure}[H]
\includegraphics[width=13.5 cm]{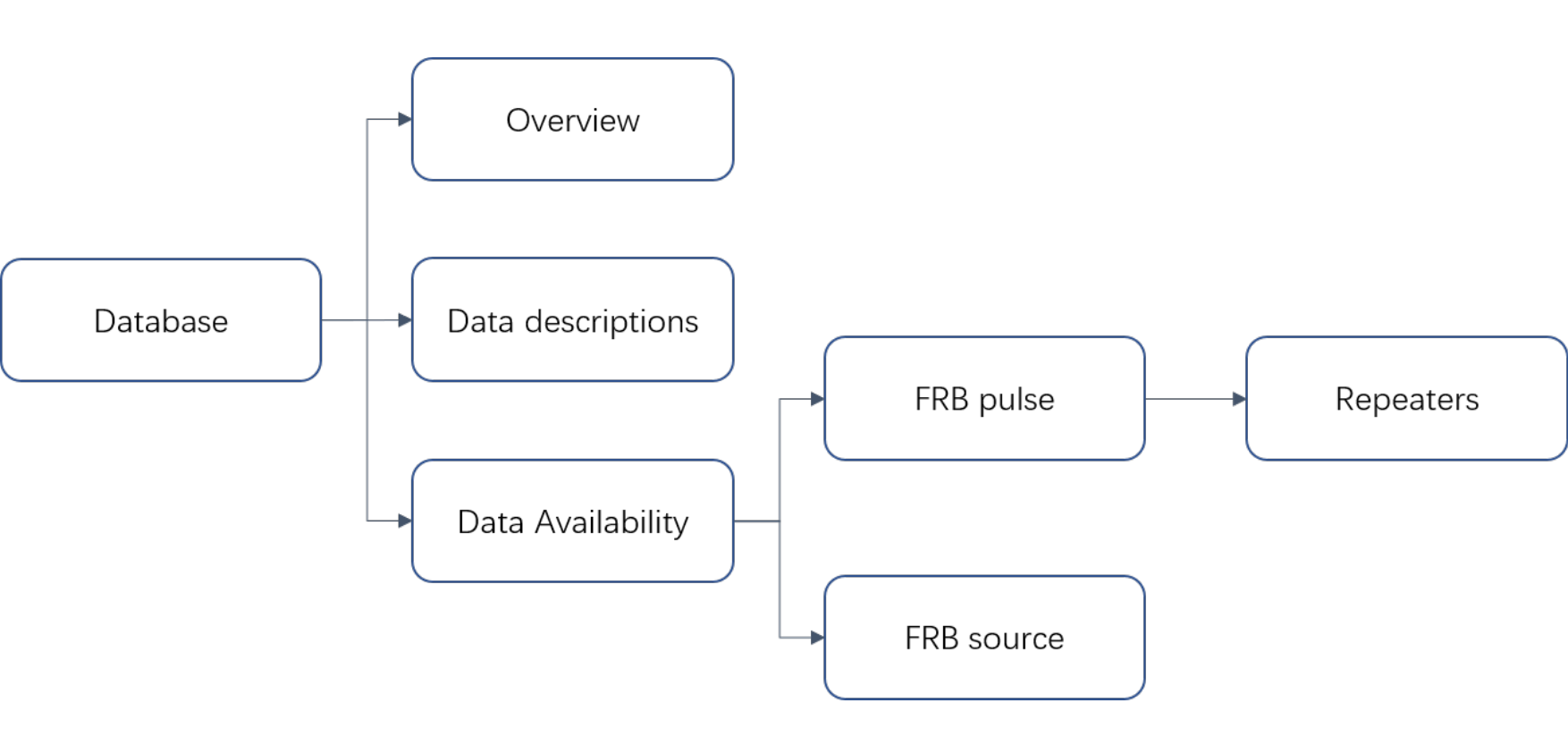}
\caption{Architecture of the 
 Blinkverse platform.\label{fig1}}
\end{figure}

MongoDB, a multi-cloud database service, offers a suitable NoSQL database to serve as the catalogue infrastructure for the platform. This platform is separated into three modules (Figure \ref{fig1}): overview, data description, and data availability. The various statistical charts on the homepage display an overall overview of the data from Blinkverse. The schema lists and description of the parameters are available in the data description. The display format of the data is divided into two types: FRB source information and pulse information, among which the repeated bursts of pulse information are listed separately. This database is under development and will be improved in the next 2-3 years under sufficient investigation to make it more useful.

\subsection{Overview}

\begin{figure}[H]
\includegraphics[width=13.5 cm]{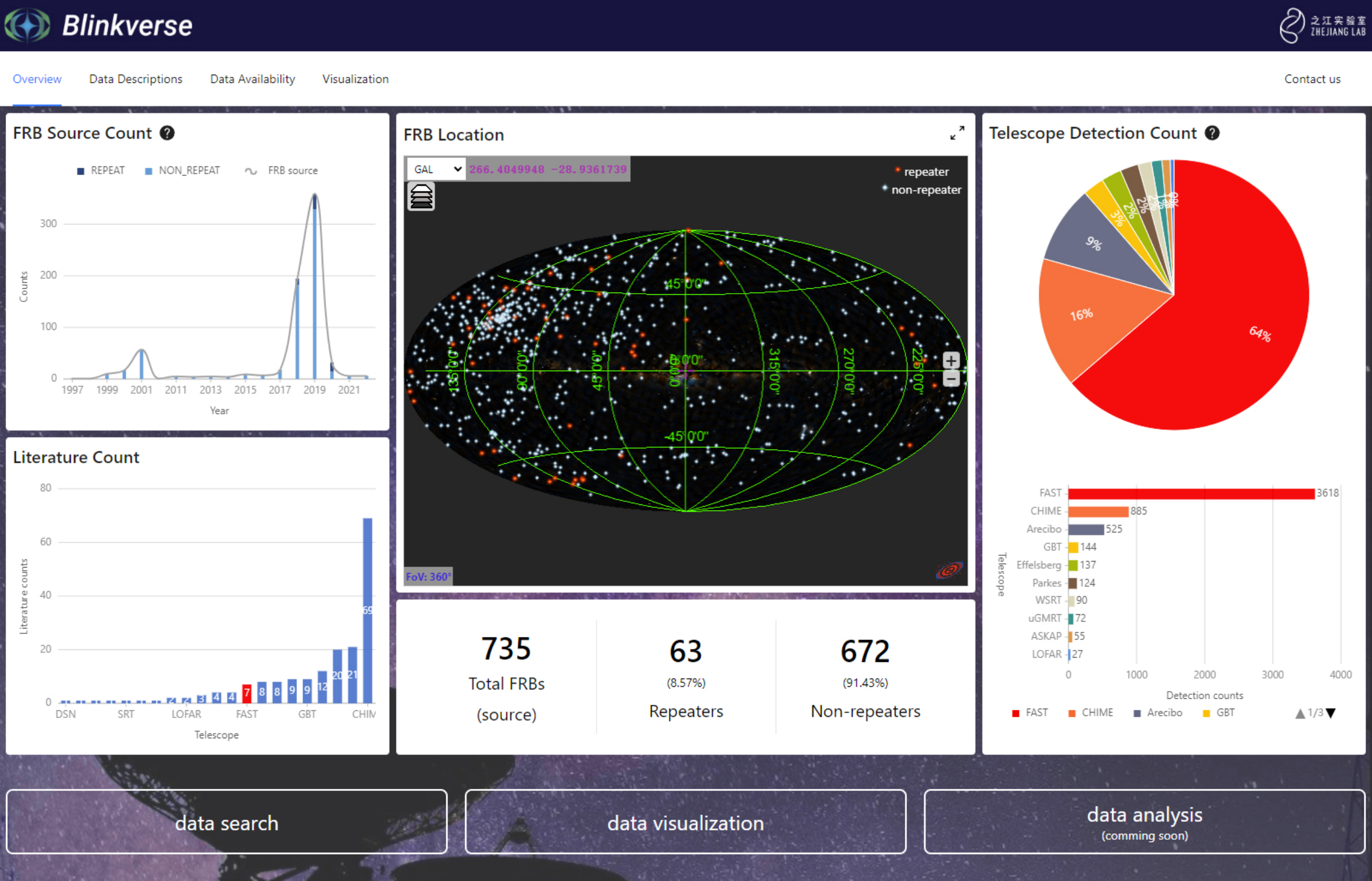}
\caption{Home page of 
 the Blinkverse platform.\label{fig2}}
\end{figure}

\begin{figure}[H]
\includegraphics[width=13.5 cm]{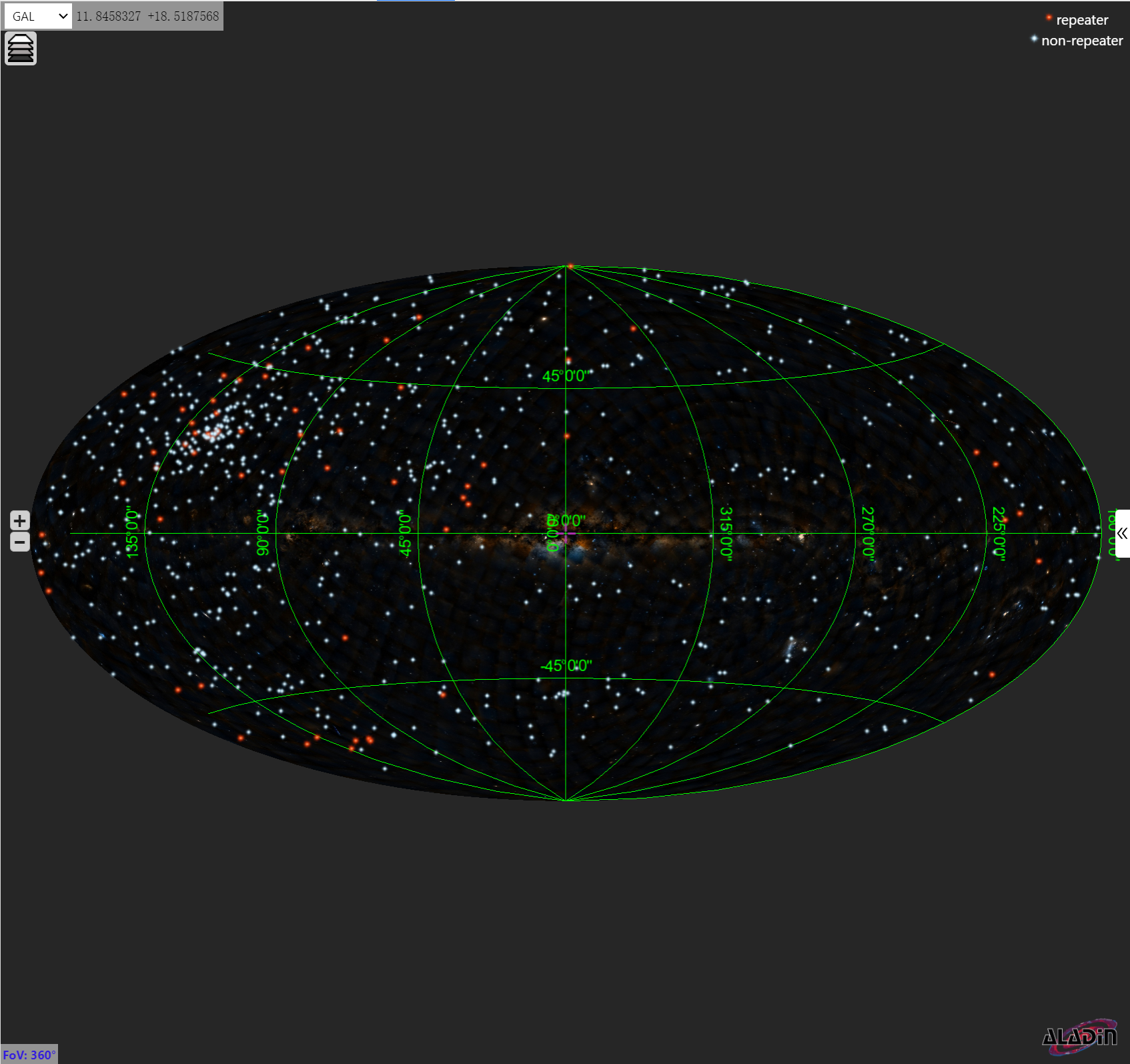}
\caption{Visualization page 
 of the Blinkverse platform.\label{fig3}}
\end{figure}

Figure \ref{fig2} displays a homepage with a statistical overview of the observed events. A celestial map is displayed in the middle of this page. All recorded FRBs have been marked on this map with white dots for non-repeaters and red dots for repeaters. The interactive operation allows users to click one of the FRBs on the map to obtain the information they want. An individual visualization page (Figure \ref{fig3}) has the same effect for choosing an interesting event.

The number of FRB discoveries is displayed below the celestial map. A total of \mbox{735 FRBs} covers 63 repeaters and 672 non-repeaters. Over 500 FRBs have been discovered in 400-800 MHz by the Canadian Hydrogen Intensity Mapping Experiment (CHIME) with a large collecting area and wide field of view since 2018 \citep{chime2018}, whereas most FRB discoveries before 2018 were made at Parkes radio telescope \citep{manchester2001}.

A pie chart in Figure \ref{fig2} displays the count of pulse detections of FRBs from various telescopes. The quantity of all the FRB bursts reaches up to \textasciitilde 5600, which mostly contributes to the repeater of FRB20121102A and FRB20201124A detected by the Five-hundred-meter Aperture Spherical radio Telescope (FAST) with high sensitivity \citep{li2021,xu2022}. 

\subsection{Data Description and Data Availability}

We reviewed multiple observational papers and relevant database websites to identify common parameters used for characterizing the properties of FRBs \citep{spitler2016,chime2019a,chime2019b,li2021,hewitt2022,xu2022}. All the data in the database were obtained from various studies in the literature and datasets. The relevant links to the references for each burst are provided on Blinkverse. Based on our findings, we proposed new schema lists (see Table \ref{tab1}) with improved descriptions of various aspects of FRBs. Two types of schema list have been created to record the information on the burst properties and positions of FRB sources. We may add or modify parameters if necessary in the future.

\begin{figure}[H]
    \includegraphics[width=13.5 cm]{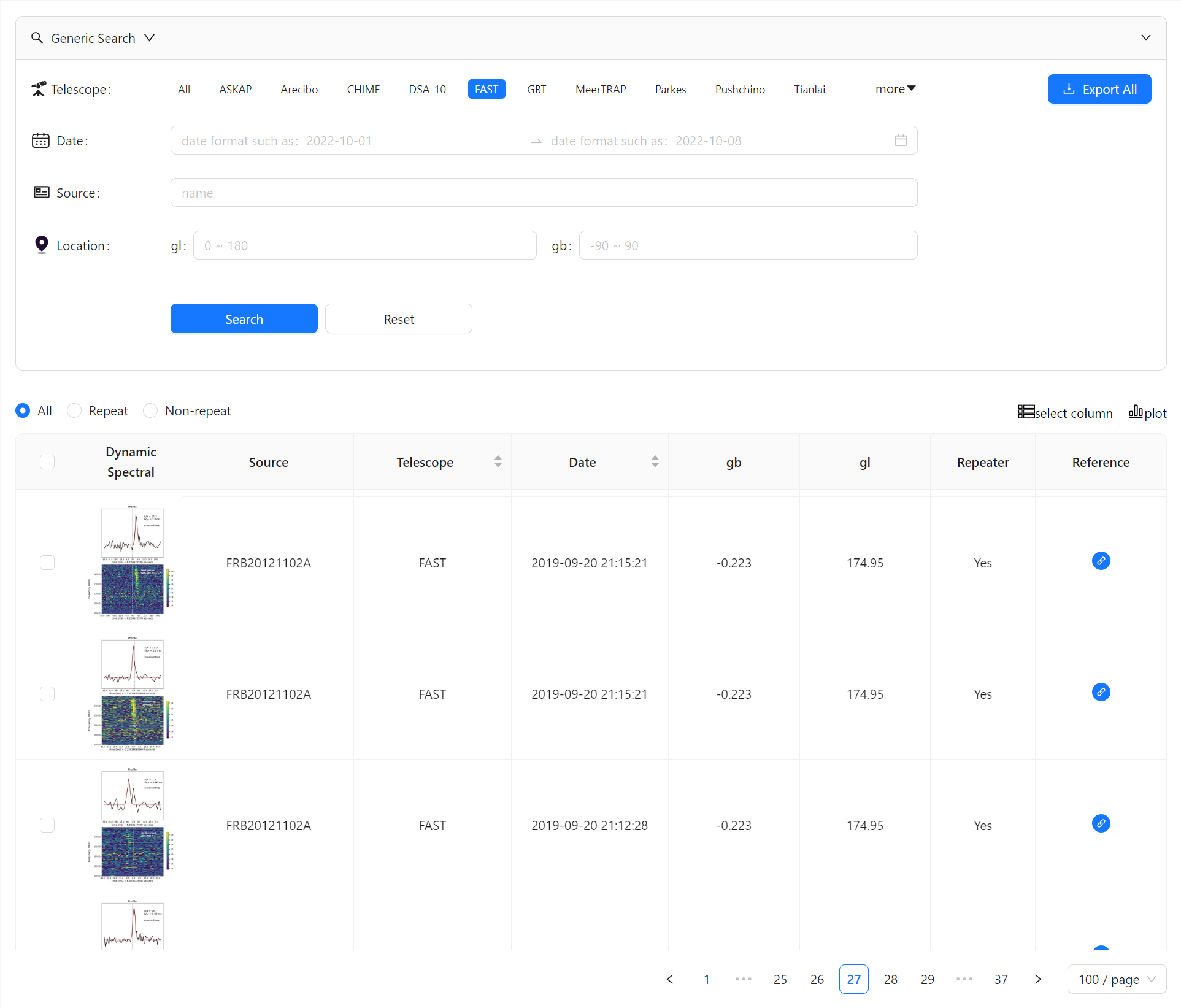}
    \caption{Data availability 
 of the Blinkverse platform.\label{fig4}}   
\end{figure}

Figure \ref{fig4} shows the generic search options and the portion of the FRB properties. The generic search for FRB sources includes telescope, observational date, FRB name, or position. In addition, we also provide an advanced search that supports logical relationship statements for convenient searching of specified parameters or a combination of parameters. Based on the already searched FRB sources, users can further select the desired parameters to download. The way to obtain data from the database is simple and flexible. We provide a download button on the website. Users can choose the parameters we provided (see Table \ref{tab1}) and click the download button to obtain the data in CSV format. Additionally, we also provide an online mapping service, where choosing the parameters for the x-axis and y-axis can facilitate drawing curves or scatterplots online.

The burst properties and positions of FRB sources are restored in the database using the name of the FRB (for example, ``FRB20121102A'') as a connector. The name of each FRB has been marked with a label of ``REPEAT'' or ``NON\_REPEAT'' in the database to distinguish between repeaters and non-repeaters. A separate page is designed for repeaters due to their significance in research. Users can click on a repeater FRB and see all of its individual bursts and dynamic spectra.

\vspace{-6pt}

\begin{table}[H]
\caption{Comparison of data and services of main data websites.\label{tab1}}
	\begin{adjustwidth}{-\extralength}{0cm}
		\begin{tabularx}{\fulllength}{m{0.2\textwidth}<{\raggedright}m{0.2\textwidth}<{\raggedright}m{0.6\textwidth}<{\raggedright}}
			\toprule
			Parameter & Units & Description \\
			\midrule
            {Burst Properties} &\\
            \midrule
            ID &  --  & Identifier of each burst \\
            Source &  --  & Name of the FRB in TNS format \\
            Telescope &  --  & Name of the telescope \\
            Receiver &  --  & Receiver mounted on the telescope \\
            Observing\_band &  MHz  & Range of observation frequency \\  
            MJD \textsuperscript{1}  &  --   & Modified Julian Date \\ 
            SNR           &  --  & Signal to Noise \\ 
            DM\_snr       &  pc cm$^{-3}$  & Dispersion measure obtained by maximizing S/N             \\ 
            DM\_alig      &  pc cm$^{-3}$  & Dispersion measure obtained by the best burst alignment   \\ 
            Flux\_density &  Jy          & Peak flux density                                         \\ 
            Width         &  ms          & Width                                                     \\
            Freq\_c       &  MHz         & Center frequency of the observation                       \\
            Freq\_low     &  MHz         & Lowest frequency of the observation                       \\
            Freq\_up      &  MHz         & Highest frequency of the observation                      \\ 
            Fluence       &  Jy ms       & Fluence                                                   \\ 
            Energy \textsuperscript{2}   &  erg  & Energy                                \\ 
            Polar\_l      &  percent  & Fractional linear polarization                            \\ 
            Polar\_c      &  percent  & Fractional circular polarization                          \\ 
            RM\_syn \textsuperscript{3}    &  rad m$^{-2}$  & Rotation measure obtained by RM synthesis \citep{burn1966}\\ 
            RM\_qufit \textsuperscript{3}  &  rad m$^{-2}$  & Rotation measure obtained by QUFIT \citep{osullivan2012}  \\
            Scatt\_t      &  ms   & Scattering timescale                                   \\
            Scin\_f       &  MHz  & Scintillation bandwidth                                   \\
            \midrule
            {Source information 
}  & \\ 
            \midrule
            RA            &  degrees  & Right ascension in J2000 coordinates                      \\ 
            DEC           &  degrees  & Declination in J2000 coordinates                          \\
            Gal. Long.    &  degrees  & Galactic longitude                                        \\
            Gal. Lat.     &  degrees  & Galactic latitude                                         \\ 
            Repeater      &  --  & Identifier of repeater                                    \\
            DM\_ne2001 \textsuperscript{4}  &  pc cm$^{-3}$ & DM determined by NE2001 \citep{cordes2002}                \\ 
            DM\_ymw16 \textsuperscript{4}   &  pc cm$^{-3}$  & DM determined by YMW16 \citep{yao2017}                    \\
            Reference     &  --  & URL of the burst discovery paper where the event was \mbox{first reported}\\
			\bottomrule
		\end{tabularx}
	\end{adjustwidth}
\noindent{\footnotesize{\textsuperscript{1} MJD is corrected to the solar system barycenter and referenced to infinite frequency. Considering the fact that the arrival time of the pulse is influenced by the motion of the Earth, the arrival times of the pulse are transformed  to the solar system barycenter using the software \texttt{pintbary} \cite{luo2021}.}}{\footnotesize{\textsuperscript{2} The unit of energy is $10^{37}$ erg.}} {\footnotesize{\textsuperscript{3} RM Synthesis and RM-QUfitting were developed by Burn et al. (1966)\citep{burn1966} 
 and osullivan et al. (2012)\citep{osullivan2012}. We just record the data from references without any modification. The value of RM is empty when these parameters are absent in the literature.}} {\footnotesize{\textsuperscript{4} DM\_ne2001 and DM\_ymw16 were developed by Cordes et al. (2002)\citep{cordes2002} and Yao et al. (2017)\citep{yao2017}. We just record the data from references without any modification. The value of DM is empty when these parameters are absent in the literature.}}
\end{table}

\subsection{Comparison with Other Data Websites}

Table \ref{tab2} compares our Blinkverse platform with other main data websites. The Blinkverse database is a comprehensive platform and includes information from multiple observation devices, multiple bands, FRB host galaxies, corresponding dynamic spectrum charts, diverse visualization, and a simplified interface. An explanation of Table \ref{tab2} is provided below:

Telescope: The databases in Table \ref{tab2}, except CHIME/FRB, contain a large number of data obtained by various telescopes. CHIME/FRB is a special database that only preserves the FRB data obtained by the CHIME telescope. 

Host galaxy: All the databases record ``ra'' and ``dec'' to describe the position of an FRB. FRBCAT calculates and records ``redshift'' in addition. 

Dynamic spectra: We provide an interactive interface to show the dynamic spectra of FRB bursts. For a specific FRB source, the burst spectrum from every different epoch can be readily queried and presented. This offers users a much more readable data visualization platform, and as a consequence, the user will be able to identify each spectrum easily and conduct more efficient data analysis. This feature surpasses other databases or studies in the literature where all the spectra are usually only presented on a single collective figure. 

Search: We provide the generic search for FRB sources including telescope, observational date, FRB name, or position, and the advanced search that supports logical relationship statements. Conversely, only FRB names can be searched in CHIME and FRBCAT. 

Visualization: TNS, FRBCAT, FRBSTATS, and CHIME/FRB only provide lists of FRB bursts without any visualization. Blinkverse has an interactive visualization interface. The positions of FRBs are marked on the celestial sphere. 

Update: The frequency of the database update is not regular according to our experience. In contrast, Blinkverse is updated regularly every week. 

Download: Download formats supported by the database. 

\begin{table}[H]
\caption{Comparison of data and services of main data websites.\label{tab2}}
	\begin{adjustwidth}{-\extralength}{0cm}
		\newcolumntype{C}{>{\centering\arraybackslash}X}
		\begin{tabularx}{\fulllength}{l*3{c}*4{c}}
			\toprule
			\multicolumn{1}{c}{ } & \multicolumn{3}{c}{\textbf{Data}} & \multicolumn{4}{c}{\textbf{Services}} \\
			\midrule
            {} & \textbf{Telescope} & \textbf{Host Galaxy} & \textbf{Dynamic Spectra} & \textbf{Search} & \textbf{Visualization} & \textbf{Update} & \textbf{Download} \\
            \midrule
            TNS         & multiple   & ra, dec     & unedited        & complex  & lack    & normal  & CSV, TSV          \\
            FRBCAT      & multiple   & ra, dec, z  & lack            & single   & lack    & stopped & CSV, FITS         \\
            FRBSTATS    & multiple   & ra, dec     & lack            & lack     & lack    & normal  & CSV, JASON        \\
            CHIME/FRB   & unique     & ra, dec     & partial         & single   & lack    & normal  & CSV, FITS, JASON  \\ 
            Blinkverse  & multiple   & ra, dec     & partial, edited & diverse  & diverse & normal  & CSV               \\
			\bottomrule
		\end{tabularx}
	\end{adjustwidth}
\end{table}

\section{Examples of Data Mining with Blinkverse}\label{sec3}

Users can easily access data from the Blinkverse database via the REST API, which is an API that conforms to the design principles of REST, using the \texttt{requests} module. The well-defined data structure enables straightforward data analysis. Users can download the data from the website and read it into a \texttt{DataFrame} format using \texttt{pandas}, or directly access \texttt{DataFrame} format data by calling the API using the provided sample code. Here, we present several simple examples of data analysis.

\begin{figure}[H]
    \includegraphics[width=0.6\textwidth]{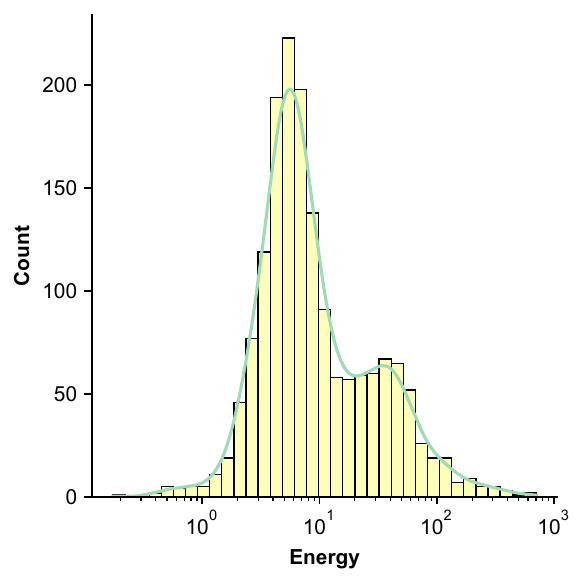}
    \caption{Energy distribution of bursts from FRB~20121102 since MJD = 58724. The green line denotes the kernel density estimate (KDE) of the energies.}
    \label{fig:energy}
\end{figure}

Upon obtaining the data, the first step is to check their distribution. Taking energy as an example, we replicated the energy distribution shown in \cite{li2021} for FRB~20121102A using \texttt{seaborn.displot}. By filtering out bursts from source FRB~20121102A and MJD > 58724, we can show the bimodal energy distribution, as in Figure~\ref{fig:energy}.

\begin{figure}[H]
    \includegraphics[width=0.6\textwidth]{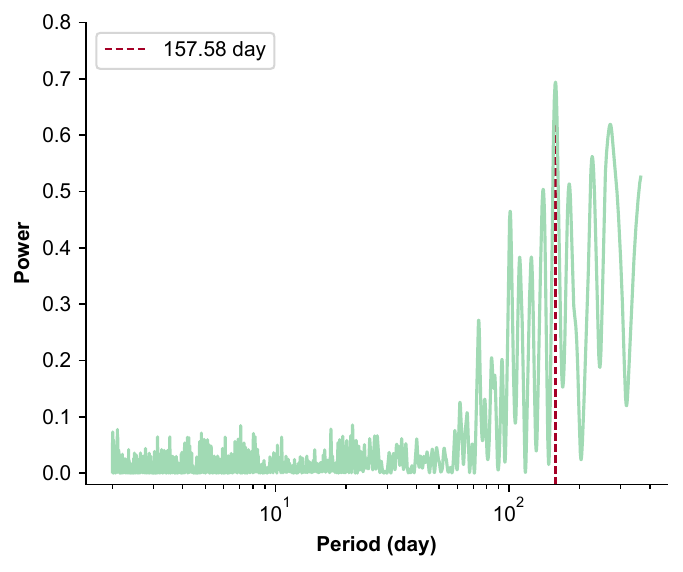}
    \caption{Lomb--Scargle periodogram of FRB~20121102A. The red line depicts the position of the period corresponding to the maximum power, which is approximately 157 days.}
    \label{fig:period}
\end{figure}

Furthermore, the Blinkverse database contains relatively complete and long-term burst information, making the search for long FRB periods possible and easy. Similarly, we used FRB~20121102A as an example to search for its long period. We extracted the MJD column from the data filtered by source FRB~20121102A, and used \texttt{scipy.signal.lombscargle} to calculate the power of the period in the range of 2--365 days. In Figure~\ref{fig:period}, we reproduce the 157-day period of FRB~20121102A \citep{rajwad2020, cruces2020}.

\vspace{-9pt}

The Blinkverse database records various properties for bursts, making multi-parameter analysis or FRB classification possible. We selected bursts having DM, Flux, Fluence, Width, and Freq, and attempted to classify FRBs using these five parameters.

Here, we used two methods, decision trees and random forests, to show the classification of FRB. Decision trees are a supervised learning method that uses a tree-like structure to represent decision rules to solve classification problems \citep{breiman1984}. Random forests are an ensemble learning algorithm composed of multiple decision trees. Their basic idea is to construct different decision trees by randomly selecting samples and features, and then vote or average the classification results of each tree to obtain the final prediction \citep{breiman2001}. Random forests have high accuracy and generalization performance.

\begin{figure}[H]
    \includegraphics[width=0.6\textwidth]{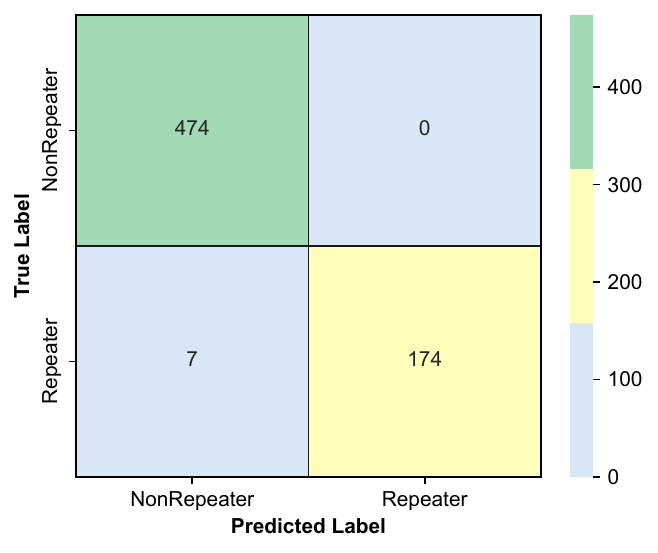}\\
    \includegraphics[width=0.6\textwidth]{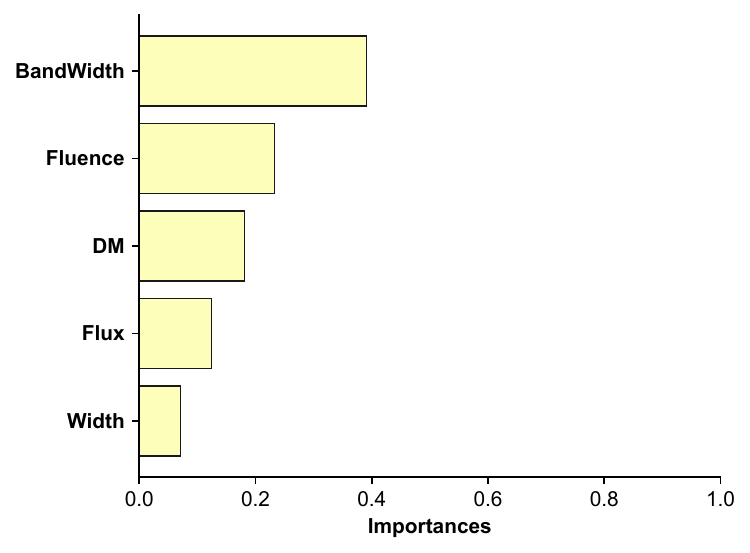}
    \caption{\textbf{Top panel}: the confusion matrix of the random forest model used for FRB classification, demonstrating that the majority of the bursts were correctly categorized. \textbf{Bottom panel}: the relative importance of the parameters used for classification.}
    \label{fig:rf}
\end{figure}

The confusion matrix is a table used to evaluate the performance of a classification model, showing the number of correct and incorrect predictions of the model for each class. The confusion matrix after fitting the data with random forests is shown in Figure~\ref{fig:rf}, indicating that only a small number of bursts are misclassified. The majority of bursts can be correctly predicted by the model. In addition, by examining the importance of parameters in the random forest model, it can be seen that \textit{Bandwidth} and \textit{Fluence} contribute the most to FRB classification. This is consistent with previous research, indicating that non-repeating FRBs typically are brighter and have wider bandwidth than repeating \mbox{FRBs \citep{chime2021, chenhy2022, chenbh2022, luo2023, zhu2023}}.

\begin{figure}[H]
    \includegraphics[width=0.6\textwidth]{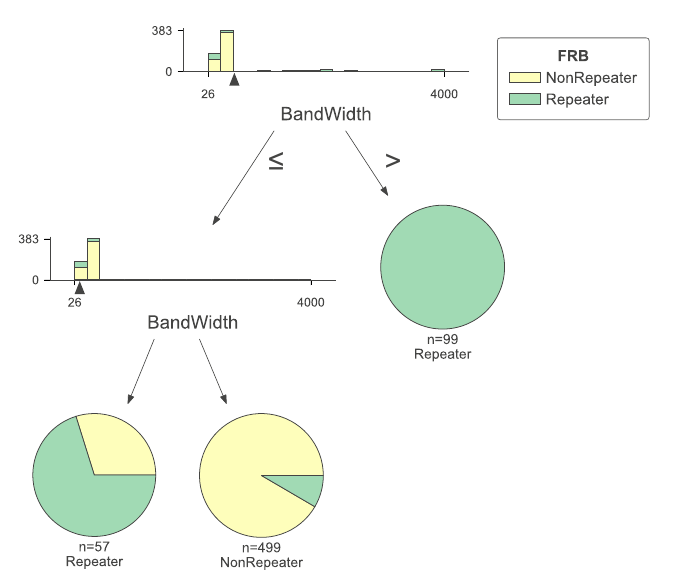}
    \caption{The decision path of decision trees used for classifying FRBs.}
    \label{fig:dt}
\end{figure}

As decision trees are prone to overfitting, we only used a two-level decision tree to classify FRBs (Figure~\ref{fig:dt}), and similarly, we found that \textit{Bandwidth} was the most important parameter for distinguishing between repeating and non-repeating FRBs.

\begin{figure}[H]
    \includegraphics[width=0.6\textwidth]{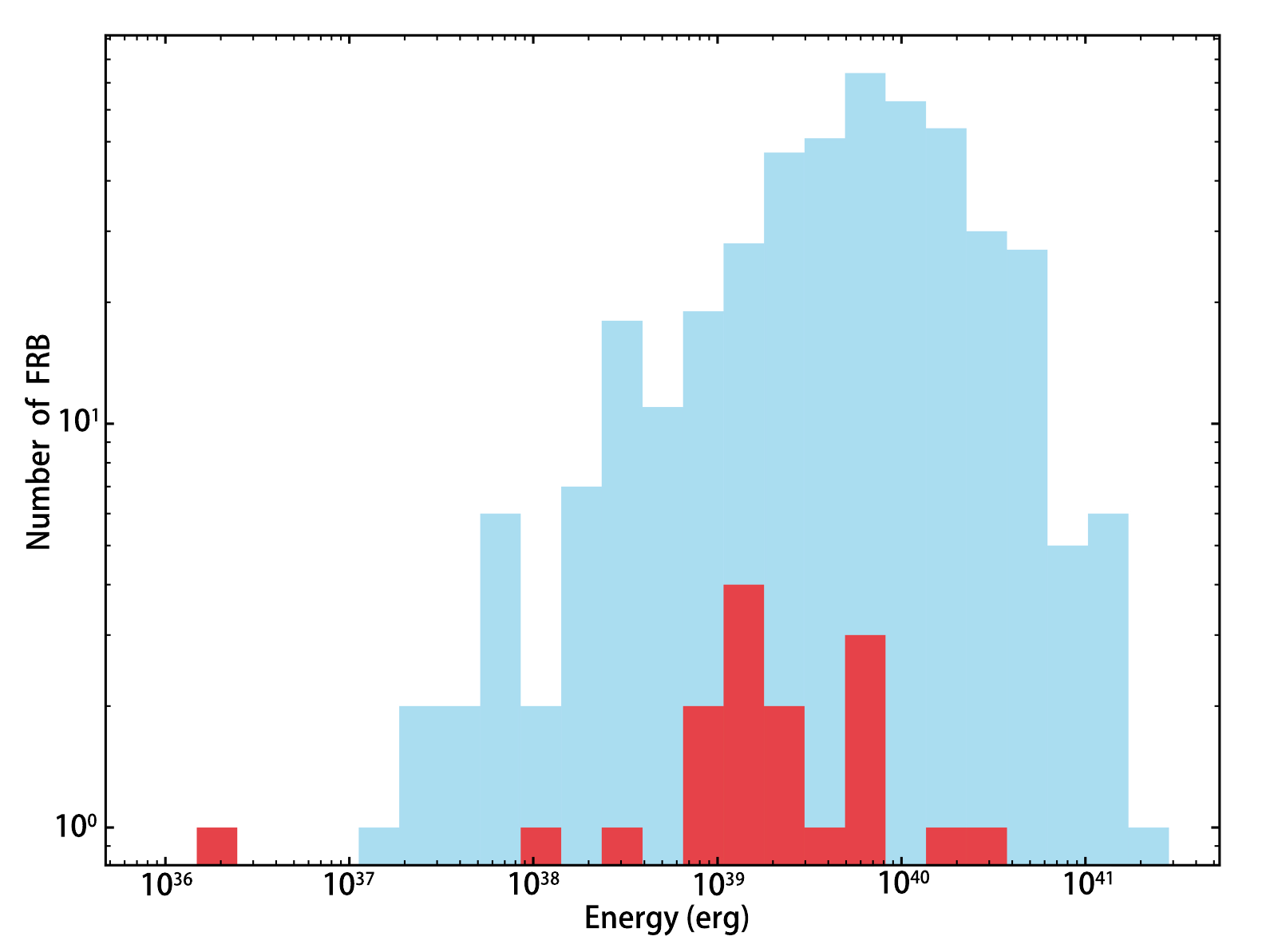}
    \caption{Energy distribution of repeaters in red bars and non-repeaters in blue bars.}
    \label{fig:hist}
\end{figure}

We calculated the values of energy to classify the repeaters and non-repeaters. The isotropic equivalent burst energy is calculated following the equation

\[E = 10^{39 } ~\mathrm{erg} \frac{4\pi}{1+z} \left( \frac{D}{10^{28}~\mathrm{cm}} \right) \left(\frac{F}{\mathrm{Jy\cdot ms}} \right)\left( \frac{\nu}{\mathrm{GHz}}\right),\]
where \(z\) is the redshift, if the redshift is measured by the emission lines detected in the high-S/N LRIS spectrum, the $z$-value is used to calculate the luminosity distance (\(D\)) adopting the standard \emph{Planck} cosmological model \citep{planck2020}. If the redshift is not measured, the distance and redshift can be calculated using the YMW16 electron density model \citep{yao2017}; \(F = S_{\nu} × W_{\mathrm{eq}}\) is the specific fluence, \(S_{\nu}\) is the peak specific flux, and \(\nu\) is the observed frequency of each pulse. We calculated the energy distributions of repeated and non-repeated bursts separately, as shown in Figure \ref{fig:hist}. Using the K-S test, we obtained a \emph{p}-value of 0.0097, which is less than 0.05, indicating that the distributions of the two groups are different. Consistent with CHIME observations, repeater bursts have a longer duration and are narrower in bandwidth than non-repeater bursts \citep{pleunis2021}. The differences between the two groups can be verified by \mbox{several parameters.}

\section{Conclusions}\label{sec4}
We have developed a comprehensive open-access FRB database named Blinkverse. The main characteristics of Blinkverse include the following:

(1) Blinkverse has 30 parameters, such as fluence, frequency, energy, polarization, etc. (see Table~\ref{tab1}), which are more comprehensive than those in FRBCAT. 

(2) Blinkverse has an interactive visualization interface that TNS, CHIME/FRB, and FRBSTATS do not have. The positions of FRBs are marked on the celestial sphere. Users can click on the map to obtain sources and their parameters. 

(3) FRB sources can be retrieved through Blinkverse based on parameter searches and their logical combinations, making it more versatile and accessible than TNS. 

(4) Blinkverse is updated weekly. 

(5) Blinkverse facilitates the systematic analysis of the FRB population and its multi-parameter characteristics. As an example, we utilized Blinkverse to find that the energy distributions of repeaters and single events are distinct from each other.  

\vspace{6pt} 



\authorcontributions{Conceptualization, 
 D.L., C.-W.T. and Y.F.; methodology, H.C.; software, Z.K., H.W. and J.H.; validation, P.W.; formal analysis, Y.Z. (Yongkun Zhang) and J.X. (Jintao Xie); investigation, J.X. (Jiaying Xu) and Y.Z. (Yun Zheng); resources, J.X. (Jiaying Xu); data curation, H.W. and D.Z.; writing---original draft preparation, J.X. (Jiaying Xu), Y.Z. (Yongkun Zhang)m and J.X. (Jintao Xie); writing---review and editing, J.X. (Jiaying Xu) and Y.F.; visualization, J.H.; supervision, H.C. and Z.K.; project administration, D.L. and J.X. (Jiaying Xu); funding acquisition, J.X. (Jiaying Xu). All authors have read and agreed to the published version of the manuscript.}

\funding{\textls[-15]{This research is funded by National Natural Science Foundation of China grant No.\ 11988101, 12203045, Special funding from China Postdoctoral Science Foundation grant} no. 2022TQ0313, and Key Research Project of Zhejiang Laboratory no.\ 2021PE0AC03.}

\institutionalreview{Not applicable.}

\informedconsent{Not applicable.}

\dataavailability{The data presented in this study are openly available in major database sets such as 
 \url{https://www.chime-frb.ca/} for CHIME/FRB and \url{https://www.wis-tns.org/} for reported FRBs and multiple observational papers. The relevant links to the references of each burst are provided on Blinkverse.} 

\acknowledgments{This work was performed on the alkaidos at Zhejiang Laboratory. We thank all researchers who have contributed to the study of fast radio burst observations.}

\conflictsofinterest{The authors declare no conflicts of interest.} 



\abbreviations{Abbreviations}{
The following abbreviations are used in this manuscript:\\

\noindent 
\begin{tabular}{@{}ll}
MDPI & Multidisciplinary Digital Publishing Institute\\
FRB & fast radio burst\\
TNS & Transient Name Server\\
ATs & astronomical transients\\
FRBCAT & Fast Radio Burst Catalogue\\
CHIME &  Canadian Hydrogen Intensity Mapping Experiment\\
FAST &  Five-hundred-meter Aperture Spherical radio Telescope\\
MongoDB & a multi-cloud database service\\
REST & representational state transfer\\
API & application programming interface\\
\end{tabular}
}

\begin{adjustwidth}{-4.6cm}{0cm}
 \printendnotes[custom]
\end{adjustwidth}
\begin{adjustwidth}{-\extralength}{0cm}
\reftitle{References}

%



\PublishersNote{}
\end{adjustwidth}
\end{document}